\documentclass[12pt]{article}
\pdfoutput=1

\usepackage{epsf, amssymb}
\usepackage{cite}
\usepackage{epsfig}
\usepackage{amsmath}
\usepackage{latexsym}
\usepackage{epsf}
\usepackage[english]{babel}
\usepackage{graphicx,color}
\usepackage{url}

\usepackage[colorlinks=true, linkcolor=blue, bookmarks=true]{hyperref}
\newcommand{\arXiv}[1]{\href{http://www.arXiv.org/abs/#1}{#1}}

\makeatletter
\renewcommand\section{\@startsection {section}{1}{\z@}%
                               {-3.5ex \@plus -1ex \@minus -.2ex}%nn
                               {2.3ex \@plus.2ex}%
                               {\normalfont\large\bfseries}}
\renewcommand\subsection{\@startsection{subsection}{2}{\z@}%
                                 {-3.25ex\@plus -1ex \@minus -.2ex}%
                                 {1.5ex \@plus .2ex}%
                                 {\normalfont\bfseries}}
\makeatother

\def\IZ{\relax\ifmmode\mathchoice
{\hbox{\cmss Z\kern-.4em Z}}{\hbox{\cmss Z\kern-.4em Z}}
{\lower.9pt\hbox{\cmsss Z\kern-.4em Z}} {\lower1.2pt\hbox{\cmsss
Z\kern-.4em Z}}\else{\cmss Z\kern-.4em Z}\fi}
\def\IR{\relax{\rm I\kern-.18em R}}

\def\one{{\hbox{ 1\kern-.8mm l}}}

\newlength{\bredde}
\def\slash#1{\settowidth{\bredde}{$#1$}\ifmmode\,\raisebox{.15ex}{/}
\hspace*{-\bredde} #1\else$\,\raisebox{.15ex}{/}\hspace*{-\bredde}
#1$\fi}

\newcommand  {\Rbar} {{\mbox{\rm$\mbox{I}\!\mbox{R}$}}}

\newsavebox{\zzzbar}
\sbox{\zzzbar}
{\setlength{\unitlength}{0.9em}
\begin{picture}(0.6,0.7)
\thinlines
\put(0,0){\line(1,0){0.6}}
\put(0,0.75){\line(1,0){0.575}}
\multiput(0,0)(0.0125,0.025){30}{\rule{0.3pt}{0.3pt}}
\multiput(0.2,0)(0.0125,0.025){30}{\rule{0.3pt}{0.3pt}}
\put(0,0.75){\line(0,-1){0.15}}
\put(0.015,0.75){\line(0,-1){0.1}}
\put(0.03,0.75){\line(0,-1){0.075}}
\put(0.045,0.75){\line(0,-1){0.05}}
\put(0.05,0.75){\line(0,-1){0.025}}
\put(0.6,0){\line(0,1){0.15}}
\put(0.585,0){\line(0,1){0.1}}
\put(0.57,0){\line(0,1){0.075}}
\put(0.555,0){\line(0,1){0.05}}
\put(0.55,0){\line(0,1){0.025}}
\end{picture}}
\newcommand{\Zbar}{\mathord{\!{\usebox{\zzzbar}}}}

\newcommand{\bra}[1]{\langle{#1}|}
\newcommand{\ket}[1]{|{#1}\rangle}

\newcommand{\ena}{\end{eqnarray}}

\newcommand{\be}{\begin{equation}}
\newcommand{\ee}{\end{equation}}

\newcommand{\Tr}{{\rm Tr}}
\def\be{\begin{equation}}
\def\ee{\end{equation}}

\def\r{\rho}

\def\({\left (}
\def\){\right )}
\def\[{\left [}
\def\[{\right ]}

\def\ba{\begin{eqnarray}}
\def\ea{\end{eqnarray}}

\def \r{{\bf r}}

\input amssym.def
\input amssym.tex

\newcommand{\bbibitem}[1]{\bibitem{#1}\marginpar{#1}}
\def\Bibitem#1{\bibitem{#1}%
  \smash{\hbox to0pt{\raise1ex\hbox{\tiny[#1]}\hss}}}

\def\Label#1{\label{#1}%
  \smash{\hbox to0pt{\raise1ex\hbox{\tiny[#1]}\hss}}}
\def\noLabels{\let\Label=\label}
\def\nobbibitem{\let\bbibitem=\bibitem}
 \def\noBibitem{\let\Bibitem=\bibitem}
\def\[{\left [}
\def\]{\right ]}
\def\({\left (}
\def\){\right )}

\def\r{\rho}

\def\r2{\sqrt{2}}

\def\bra{{\langle}}
\def\ket{{\rangle}}
\def\Label#1{\label{#1}%
  \smash{\hbox to0pt{\raise1ex\hbox{\tiny[#1]}\hss}}}
\def\noLabels{\let\Label=\label}
\def\nobbibitem{\let\bbibitem=\bibitem}
\newcommand{\bz}[0]{\bar{z}}
\newcommand{\bx}[0]{\bar{x}}

\newcommand{\bea}{\begin{eqnarray}}
\newcommand{\eea}{\end{eqnarray}}

\newcommand{\beq} {\begin{equation}}
\newcommand{\eeq} {\end{equation}}
\newcommand{\beqa} {\begin{eqnarray}}
\newcommand{\eeqa} {\end{eqnarray}}

\newcommand{\mC}{\mathcal{C}}

\newcommand{\beqn}{\begin{eqnarray}}
\newcommand{\eeqn}{\end{eqnarray}}

\def\bra{\langle}
\def\ket{\rangle}

\usepackage{caption}
\usepackage{subcaption}
\usepackage{float}
\usepackage{braket}
\usepackage{bbm}
\usepackage{ae}

\newcommand{\Z}{\ensuremath{\mathbbm{Z}}}

\newcommand{\RF}{\right)}
\newcommand{\LF}{\left(}
\newcommand{\RT}{\right]}
\newcommand{\LT}{\left[}

\begin{document}

\begin{titlepage}
\begin{flushright}
%arXiv:
\end{flushright}
\vfill
\begin{center}
{\Large \bf  Entwinement in discretely gauged theories}

\vskip 10mm

{\large V.~Balasubramanian$^{a,b}$, A.~Bernamonti$^{c,d}$, B.~Craps$^b$,\\  T.~De Jonckheere$^b$, F.~Galli$^{c,d}$\\
\vspace{3mm}
}

\vskip 7mm

$^a$ David Rittenhouse Laboratory, Univ. of Pennsylvania, 
 Philadelphia, PA 19104, USA \\
$^b$ Theoretische Natuurkunde, Vrije Universiteit Brussel, and \\ International Solvay Institutes,
Pleinlaan 2, B-1050 Brussels, Belgium \\
$^c$ Perimeter Institute for Theoretical Physics, \\
31 Caroline Street North, ON N2L 2Y5, Canada \\
$^d$ KU Leuven, Institute for Theoretical Physics,\\
 Celestijnenlaan 200D, B-3001 Leuven, Belgium\\

\vskip 6mm
{\small\noindent  {\tt vijay@physics.upenn.edu, abernamonti@perimeterinstitute.ca, Ben.Craps@vub.ac.be, Tim.De.Jonckheere@vub.ac.be, fgalli@perimeterinstitute.ca}}

\end{center}
\vfill

\begin{center}
{\bf ABSTRACT}
\vspace{3mm}
\end{center}

We develop the notion of  ``entwinement''  to characterize the amount of quantum entanglement between internal, discretely gauged degrees of freedom in a quantum field theory.     This concept originated in the  program of reconstructing spacetime from entanglement in holographic duality.  We define entwinement formally in terms of a novel replica method which uses twist operators  charged in a representation of the discrete gauge group.   In terms of these twist operators we define a non-local, gauge-invariant object whose expectation value computes entwinement in a standard replica limit.  We apply our method to the computation of entwinement in symmetric orbifold conformal field theories in 1+1 dimensions, which have an $S_N$ gauging.  Such a theory appears in the weak coupling limit of the D1-D5 string theory which is dual to AdS$_3$ at strong coupling.   In this context, we show how  certain kinds of entwinement measure the lengths, in units of the AdS scale, of non-minimal geodesics present in certain excited states of the system which are gravitationally described as conical defects and the $M=0$ BTZ  black hole.    The possible types of entwinement that can be computed define a very large new class of quantities characterizing the fine structure of quantum wavefunctions.   

\end{titlepage}

\tableofcontents

%%%%%%%%%%%%%%%%%%%%%%%%%%%%%%%%%%%%%%%%%%%%%%%%%%%%%%%%%%%%%%%%%%%%%%%
%%%%%%%%%%%%%%%%%%%%%%%%%%%%%%%%%%%%%%%%%%%%%%%%%%%%%%%%%%%%%%%%%%%%%%%

\section{Introduction}
\label{intro}

When a quantum system finds itself in a pure state $|\psi\rangle$, the entanglement between a part $A$ of the system and its complement $\bar A$ is quantified by the entanglement entropy.  In most applications, $A$ and $\bar A$ describe\ the degrees of freedom in complementary spatial regions.   In systems with localized degrees of freedom such as spin chain models or local quantum field theory,  this corresponds to a natural separation of the total Hilbert space.  However, it is also fruitful to consider other ways of separating the Hilbert space. 

In \cite{Balasubramanian:2011wt,Agon:2014uxa}, the degrees of freedom of a local quantum field theory were separated into high and low spatial momentum modes. It was demonstrated that in a generic interacting field theory, even in the vacuum state, the long wavelength (low energy) degrees of freedom necessarily find themselves in a nontrivial reduced density matrix because of entanglement with the short wavelength (high energy) degrees of freedom. This gives rise to finite entanglement entropy, which was computed explicitly in perturbative scalar field theories. The more traditional way of describing the low energy degrees of freedom is Wilsonian renormalization. In this language, the vacuum state of a Wilsonian low energy theory is necessarily a density matrix with finite entropy  \cite{Balasubramanian:2011wt,Agon:2014uxa}. This phenomenon of UV-IR entanglement in quantum field theories could be particularly important in theories of gravity (in which ultraviolet and infrared degrees of freedom are known to couple in nontrivial ways), as well as in the ground states of strongly correlated electronic systems (see, e.g. \cite{Lundgren:2014qua}).

This also raises the question whether it is fruitful to consider other non-spatial ways of dividing the degrees of freedom of quantum field theories.  One interesting way to separate energy scales is by imagining a collection of local observers who have a finite duration $T$ over which they can make measurements.    This is a natural situation to consider, as it describes the practical constraints of most measurements.
Intuitively, given Heisenberg's energy-time uncertainty relation, such observers will be insensitive to energies smaller than the inverse duration of the experiment, so that those low energy degrees of freedom are effectively traced out, turning the accessible part of the state into a nontrivial density matrix.   Thinking in this way,  \cite{Balasubramanian:2013rqa,Balasubramanian:2013lsa,Myers:2014jia,Headrick:2014eia} proposed a new information theoretic quantity, the {\it differential entropy},  as a measure of UV-IR entanglement, at least for two-dimensional theories.  

These effects should become stronger for theories  with an energy gap that is smaller than the inverse spatial size of the system.  For example, consider relativistic theories where  the local degrees of freedom are matrices, e.g. $SU(N)$ gauge theories.  In such systems the energy gap can be much smaller than the inverse spatial size of the system so that even a set of observers with enough time to observe the entire spatial domain will not have access to the lowest energy excitations.    Another example which is easier to visualize is a  ``long string'' theory, where strings or spin chains  are multiply wound around a spatial circle, allowing for excitation wavelengths  that exceed the system size.   In both these examples,  the key to the physics lies in entanglement between ``internal" degrees of freedom (matrix components, or strands of string) that are not spatially organized.

One way to study the entanglement of gauge degrees of freedom in an $SU(N)$ theory is to break the gauge group into $SU(m)\times SU(N-m)$ while allowing for interactions between the two sectors. This could  be realized holographically by separating a stack of $N$ branes into a stack of $m$ and one of $N-m$ branes and studying entangling surfaces in the AdS$_d\times S^{10-d}$ geometry which arises in the low energy limit. Such a set-up was first considered in~\cite{Mollabashi:2014qfa} and later refined in~\cite{Karch:2014pma,Taylor:2015kda,Mozaffar:2015bda}. The authors of~\cite{Karch:2014pma,Taylor:2015kda} also considered global symmetries and in case of an $SO(11-d)$ global symmetry they proposed a quantity in the field theory which would holographically be represented by the area of caps on the internal $S^{10-d}$.

One important complication that we have glossed over so far is that in systems exhibiting gauge symmetry, even the association of degrees of freedom to spatial regions is subtle.  For example, some of the fundamental degrees of freedom, such as Wilson loops, are not local in space, making it more complicated to split up the Hilbert space according to spatial regions. Interesting work on how to define entanglement entropy in gauge theories has recently appeared (see e.g.\ \cite{Buividovich:2008gq,Donnelly:2011hn,Casini:2013rba,Radicevic:2014kqa,Donnelly:2014gva,Donnelly:2014fua, Ghosh:2015iwa,Soni:2015yga,VanAcoleyen:2015ccp,Radicevic:2015sza}), but a complete understanding is still lacking.

These questions about entanglement in quantum field theory are also linked to equally deep questions about the nature of black hole horizons and  the holographic emergence of spacetime.  It was proposed in \cite{Ryu:2006bv,Ryu:2006ef,Hubeny:2007xt} that the entanglement entropy of a spatial region $A$ in the field theory is proportional to the area of the minimal surface in AdS space that ends on the boundary of $A$.   Furthermore, \cite{Balasubramanian:2013rqa,Balasubramanian:2013lsa,Myers:2014jia,Headrick:2014eia} showed that  the area of closed surfaces in the bulk of AdS can be related to a measure of UV-IR entanglement, the differential entropy discussed above, at least for two-dimensional boundary theories and higher dimensional cases with translational symmetries -- some of the limitations were discussed in \cite{Engelhardt:2015dta}.  Finally, in \cite{VanRaamsdonk:2010pw, Bianchi:2012ev, Faulkner:2013ica, Maldacena:2013xja,Lashkari:2013koa,Swingle:2014uza} it was suggested that spacetime connectedness is related to entanglement of the underlying quantum degrees of freedom, and   that the linearized equations of motion of gravity can be derived from the dynamics of entanglement perturbations. 

In general,  can all of spacetime geometry be reconstructed from spatial entanglement entropy in the AdS/CFT correspondence?   At least when we do not consider bulk quantum corrections to the entanglement entropy \cite{Faulkner:2013ana}, the answer is no -- in some asymptotically AdS spacetimes, the minimal surfaces anchored on the boundary that geometrically reproduce the entanglement entropy will not penetrate a region \cite{Hubeny:2012ry, Engelhardt:2013tra} which has been called the {\it entanglement shadow} \cite{Balasubramanian:2014sra,Freivogel:2014lja}.    It is argued in \cite{Balasubramanian:2014sra} that in such systems entanglement can be dominated by ``internal'' degrees of freedom (e.g.\ the matrix components, or strands of string discussed above) that are not spatially organized, and that these entanglements can measure the areas of non-minimal, extremal surfaces that can penetrate part of the entanglement shadows of the gravitational dual.  In the examples arising in the AdS/CFT correspondence, such internal degrees of freedom are usually gauged.  Thus, reconstructing the emergent space in gauge/gravity duality will involve entanglement between ``internal", gauged degrees of freedom --   a notion that was named {\em entwinement}  in \cite{Balasubramanian:2014sra}.  While we will not address the question of which part of a general spacetime can be probed by extremal surfaces (see, for instance, \cite{Hubeny:2012ry, Engelhardt:2013tra,Freivogel:2014lja}), it is clear that entwinement will often allow the reconstruction of a larger part of spacetime than spatial entanglement entropy. Entwinement also plays a key role in the description of holographic spacetimes using methods of integral geometry based in kinematic space \cite{Czech:2014ppa, Czech:2015qta, Czech:2015kbp}.

In summary, both in quantum field theory and in quantum gravity, we are driven to consider a new notion of ``entwinement'' --  non-spatial quantum entanglement  between gauged degrees of freedom.   In this paper we will define entwinement formally in discretely gauged theories, and discuss how it can be explicitly computed.  Section~\ref{sec:entdef} develops the general formalism.
For two-dimensional theories, we define entwinement in terms of a replica method using twist operators that are charged under the discrete gauge group. We use these operators to construct a new non-local, gauge invariant object whose expectation value is defined to be the entwinement in a standard replica limit.  
Section~\ref{sec:orbifold} applies this formalism to symmetric orbifold conformal field theories in two dimensions.  By explicitly applying uniformization maps, obtained by generalizing a construction of \cite{Asplund:2011cq}, to the Riemann surfaces arising from the replica method we directly compute entwinement in generic microstates of these theories.  We also comment on how to recover the usual spatial entanglement entropy as a special limit of entwinement. 

In section~\ref{sec:D1D5}, we apply our formalism to the weak coupling limit of the D1-D5 CFT, a theory with a holographic dual. Ordinary spatial entanglement entropy in generic classes of the D1-D5 CFT was considered before in~\cite{Asplund:2011cq,Giusto:2014aba,Galliani:2016cai}. For generic microstates, entanglement entropy was computed approximately using both a short interval expansion~\cite{Giusto:2014aba} and large $c$ methods~\cite{Galliani:2016cai}. In~\cite{Asplund:2011cq}, a specific state corresponding to a local quench was considered, and the evolution of the entanglement entropy was computed using a uniformization map -- we generalize the construction for use with other  microstates.
 Although this uniformization map will work for computing single interval entwinement in general microstates, in section~\ref{sec:D1D5} to compare with holography we focus on two examples of microstates, which are gravitationally related to conical defects and the $M=0$ BTZ black hole, respectively. We demonstrate that the lengths of non-minimal geodesics that penetrate entanglement shadows of the spacetime are computed by certain entwinements. In the same way as spatial entanglement entropy corresponds to minimal extremal area surfaces in the bulk, here entwinement provides a direct field theory interpretation for non-minimal extremal surfaces.  
 
The paper concludes with a discussion of possible directions forward to develop the notion of entwinement in more general situations, and comments on the relation with the appearance of non-minimal geodesics in the semiclassical Virasoro conformal blocks discussed in \cite{Asplund:2014coa}.
A number of technical results are collected in appendices.

\setcounter{equation}{0}
%%%%%%%%%%%%%%%%%%%%%%%%%%%%%%%%%%%%%%%%%%%%%%%%%%%%%%%%%%%%%%%%%%%%%%%
\section{Defining entwinement}
\label{sec:entdef}

In a gauge theory, states are required to be symmetric under identifications by the gauge group.  
The main complication in discussing entanglement entropy in gauge theories is that the Hilbert space does not factorize.   For example, for a $U(1)$ gauge theory there is  a Gauss law constraint which requires that the electric flux entering a region should determine the flux leaving it.  We will be interested in situations where the gauge symmetry is discrete.  

Recently various authors have developed a formalism for dealing with this lack of factorization in gauge theories.  One method is to consider an extended Hilbert space in which the gauge constraints are temporarily relaxed \cite{Buividovich:2008gq,Donnelly:2011hn,Donnelly:2014gva,Donnelly:2014fua,Ghosh:2015iwa, Soni:2015yga, VanAcoleyen:2015ccp}.  A way of achieving this is by introducing ``edge modes'' on the entangling surface \cite{Donnelly:2011hn}. In this approach, the Hilbert space splits into superselection sectors defined by the fluxes at the entangling surface, making the reduced density matrix block-diagonal.  The entanglement entropy then becomes a sum of two contributions, a Shannon entropy associated to the distribution over superselection sections and a distillable piece arising from entanglement within each superselection sector.   In the same spirit, the approach to entwinement suggested in \cite{Balasubramanian:2014sra} was to ungauge the theory, compute, and then symmetrize.

An alternative approach is to define a subalgebra of gauge invariant observables ${\cal O}_A$  associated to the region $A$.  Then, given a density matrix $\rho$ for the full theory, the reduced density matrix $\rho_A$ is defined as the element of the subalgebra of region $A$ such that $\Tr(\rho_A {\cal O}_A) = \Tr(\rho {\cal O}_A)$ for all ${\cal O}_A$  \cite{ Casini:2013rba, Radicevic:2014kqa,Radicevic:2015sza}.   In this formalism, the reduced density matrix splits into blocks according to superselection sectors determined by the center of the subalgebra.   An algebraic approach to entwinement was recently proposed in \cite{Lin:2016fqk}, where it was shown that for a spin system the entwinement could be recovered from a state-dependent subalgebra.   

A third approach, which works for states that can be constructed by a Euclidean path integral, is the replica trick.  In this case, the entanglement entropy is computed by analytically continuing the R\'{e}nyi entropies.  In two dimensions this approach is particularly convenient and the R\'{e}nyi entropies can be defined in terms of the correlation functions of twist operators that splice together replicated copies of the CFT. At least for the case of 2d Yang-Mills theory in de Sitter space, it was verified in \cite{Donnelly:2014gva} that the replica method gives the same result as the extended Hilbert space method described above.  Below we will define entwinement along these lines.

As a working example, consider a CFT with target space $M^N/S_N$. The CFT contains $N$ sets of fields, where each set can be viewed as coordinates on one copy of the manifold $M$, together with companion fermions in case of a supersymmetric theory. The $S_N$ indicates that we identify configurations that differ by permutations of these $N$ sets of fields. This is similar to the way one treats indistinguishable particles in quantum mechanics: wavefunctions need to be appropriately symmetrized under permutations. The $S_N$ identification is really a discrete gauge symmetry.   We can gauge fix the local symmetry and think of the fields as changing continuously from point to point (i.e., each of the $N$ copies of $M$ has a continuous string embedded in it).  The theory has so-called ``twisted sectors'' in which strings are only periodic up to permutations.  A twisted sector is labeled by a conjugacy class, which is characterized by the lengths of its permutation cycles: there will be $N_m$ cycles of length $m$ such that $\sum_m N_m m = N$.  Each cycle is referred to as a ``long string'' because it can be visualized as a string winding $m$ times.  We will refer to each winding  of the long string as a ``strand''. 

The conventional spatial entanglement entropy of an angular interval of size $\alpha$ 
can be thought of as the entanglement entropy of the union of intervals of size $\alpha$ in each of the $N$ strands of the system. Following the proposal of \cite{Balasubramanian:2014sra}, we want to define entwinement as the entanglement of intervals that extend over some strands and not others. For example, one can talk about the entwinement of an interval on a single strand.  If one considers the entwinement of a union of identical intervals in each strand, then it reduces to the conventional spatial entanglement.  Because there is a gauge symmetry, we cannot invariantly specify which strand we are talking about.  But, as we will argue below, we can meaningfully talk about things like ``the entanglement of one and a half connected strands".  
 Likewise, while one cannot invariantly ask for the entanglement of a particular strand, one can ask for the entanglement of a single strand if we do not specify which one it is. 
 This invariance can be made manifest by simply averaging the computation of entanglement of a single strand over all the strands. It is worth emphasizing that this average is {\it not} the same thing as the entanglement of the union of such intervals.  Below, we will give a mathematical definition of such quantities.

%%%%%%%%%%%%
\subsection{Replica trick -- generalities}

\begin{figure}[t]
\centering 
\includegraphics[width=0.45\textwidth]{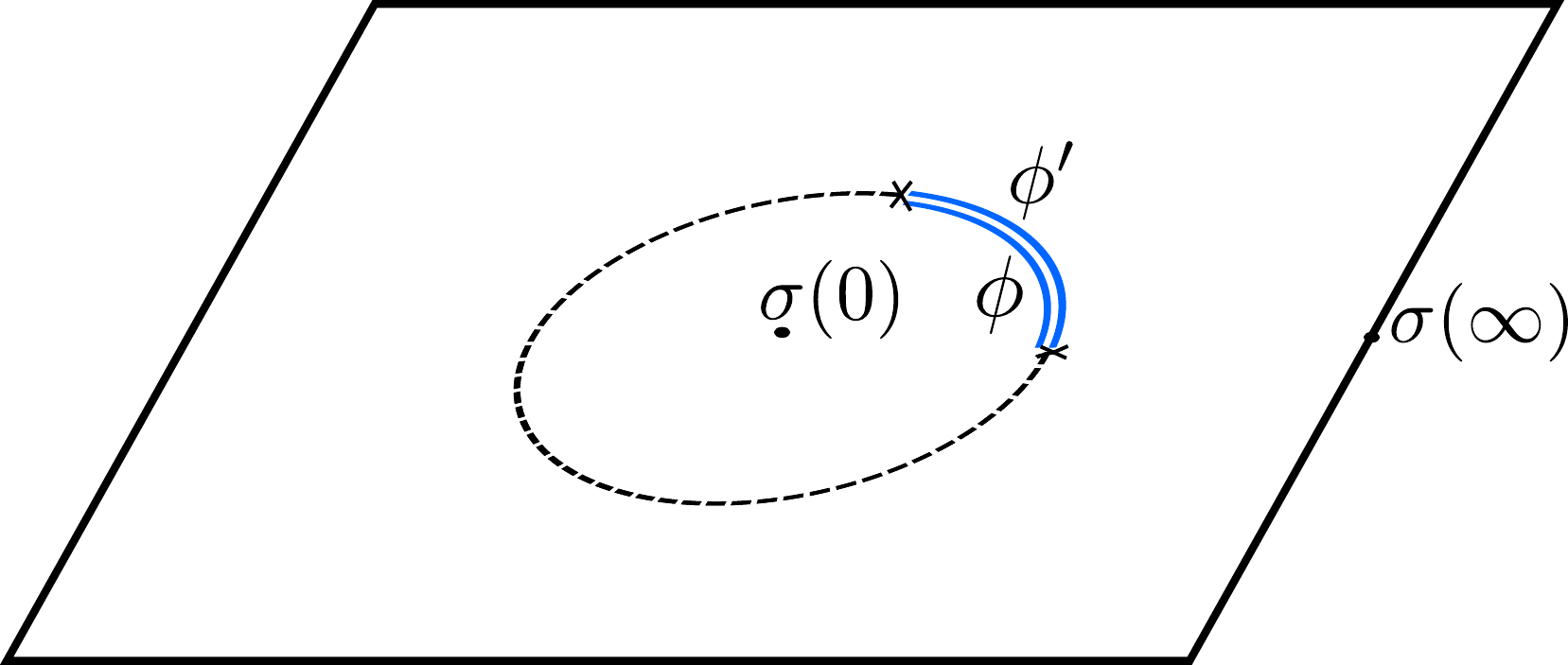}
\caption{\small  Path integral in radial quantization. The path integral is left open on the cut (indicated in full blue lines) with boundary conditions $\phi$ on the lower cut and $\phi '$ on the upper cut. The dashed lines define the complementary interval along which boundary conditions are matched $\phi=\phi '$. The operator $\sigma$ prepares the state.} \label{fig:PICut}
\end{figure}

\begin{figure}[h]
\centering 
\includegraphics[width=0.85\textwidth]{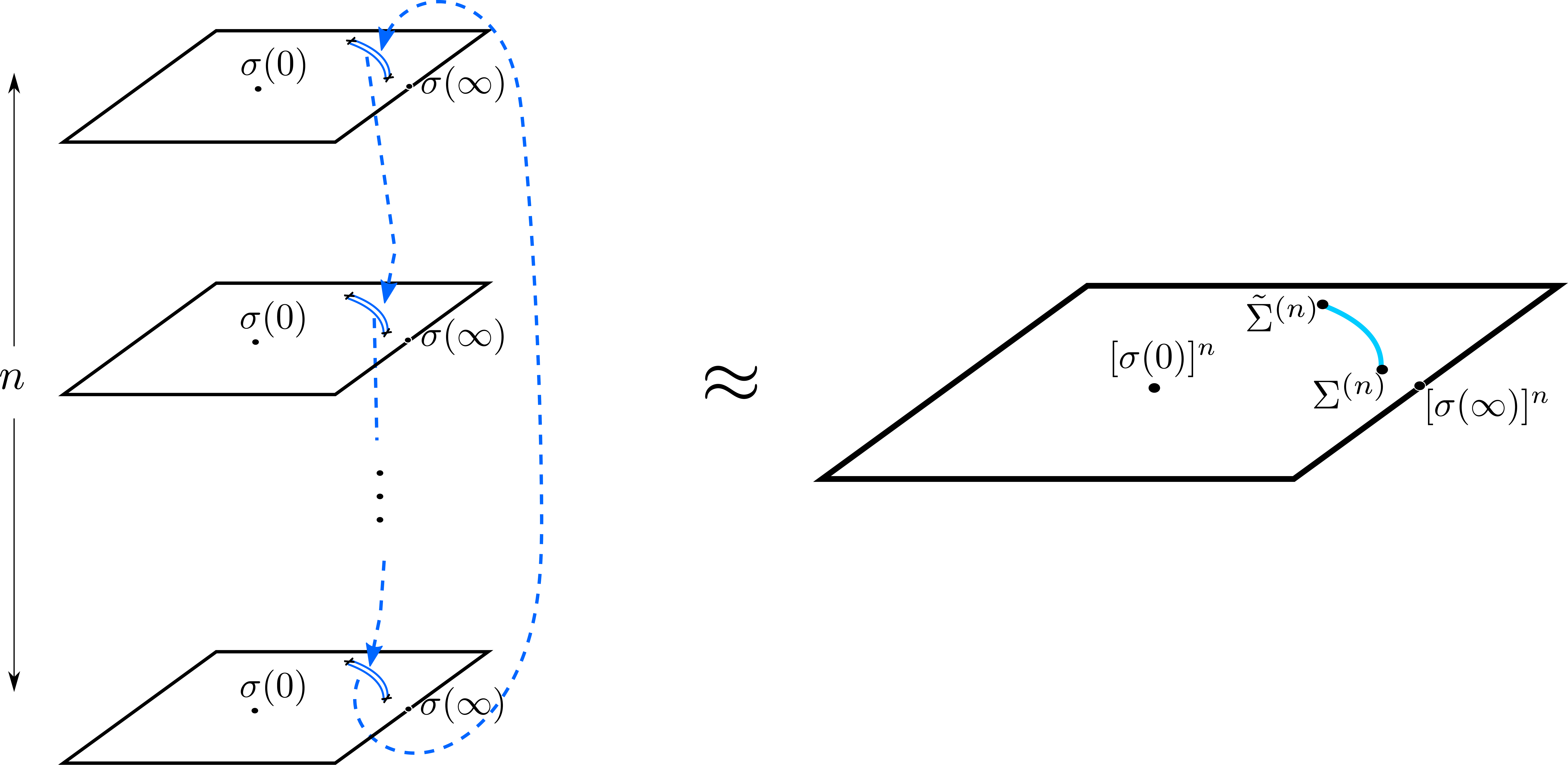}
\caption{\small {\it (Left)} The $n$-sheeted Riemann surface from cyclically gluing $n$ copies of the plane. The dashed arrows denote how to sew fields across the cuts. {\it (Right)} Correlator in the plane. The $\Sigma$ insertions represent R\'enyi twist operators, while the $\sigma$ insertions define the replicated state.} \label{fig:nsheeted}
\end{figure}

A useful method for computing entanglement entropy in two-dimensional conformal field theories is the replica trick.    In this method, the entanglement entropy of an interval $[0,\alpha]$ is computed from the reduced density matrix for this region $\rho_\alpha$ by taking a limit of the R\'{e}nyi entropies:
\begin{equation}
S(\alpha) = \lim_{n\to 1} \frac{1}{1 - n} \log \Tr(\rho_\alpha^n) \, .
\end{equation}
Consider a two-dimensional CFT in the plane in Euclidean signature.  In radial quantization,  circles of fixed radius become equal time slices.   An operator $\sigma$ acting at the origin creates a pure state, and we can find the wavefunction at $t=0$ by performing the path integral with the operator inserted and fixed boundary conditions at the unit circle.   The density matrix $\rho(\phi,\phi')$ corresponding to such a pure state is then computed by inserting operators at the origin and at infinity in the path integral, and imposing boundary conditions $\phi$ and $\phi'$ on the interior and the exterior of the unit circle, respectively.  The reduced density matrix for the interval $[0,\alpha]$ is computed by tracing over the complementary part of the unit circle (i.e., setting $\phi = \phi'$ in the complementary region and then integrating over $\phi$ there). This leaves us with a path integral over the entire plane, except over the arc corresponding to the interval $[0,\alpha]$, as depicted in figure~\ref{fig:PICut}.  
To compute $\Tr(\rho_\alpha^n)$ we consider $n$ copies of the plane cyclically glued together over the cut $[0,\alpha]$ producing an $n$-sheeted Riemann surface as in figure~\ref{fig:nsheeted} (left).   
This can alternatively be obtained as a correlator of R\'{e}nyi twist operators $\Sigma^{(n)}$ computed on a single sheet of the $n$-fold cover of the theory, ${\rm CFT}^n/{\mathbb Z}_n$, as represented in figure~\ref{fig:nsheeted} (right). The $n$ fictitious copies of the CFTs are called ``replicas".    Each of the $n$ copies of the CFT is placed in the same state.  The R\'{e}nyi twist cyclically splices together the $n$ CFT copies, such that dragging a field from one CFT around the cut produces a field in the next copy of the CFT.  

Below we will illustrate this procedure in symmetric product CFTs with target space $M^N/S_N$ and show how  entwinement in these theories can be defined in a similar manner.   We will then generalize the definition to apply to more general discretely gauged theories in two dimensions. 

\begin{figure}[h]
\centering 
\includegraphics[width=0.8\textwidth]{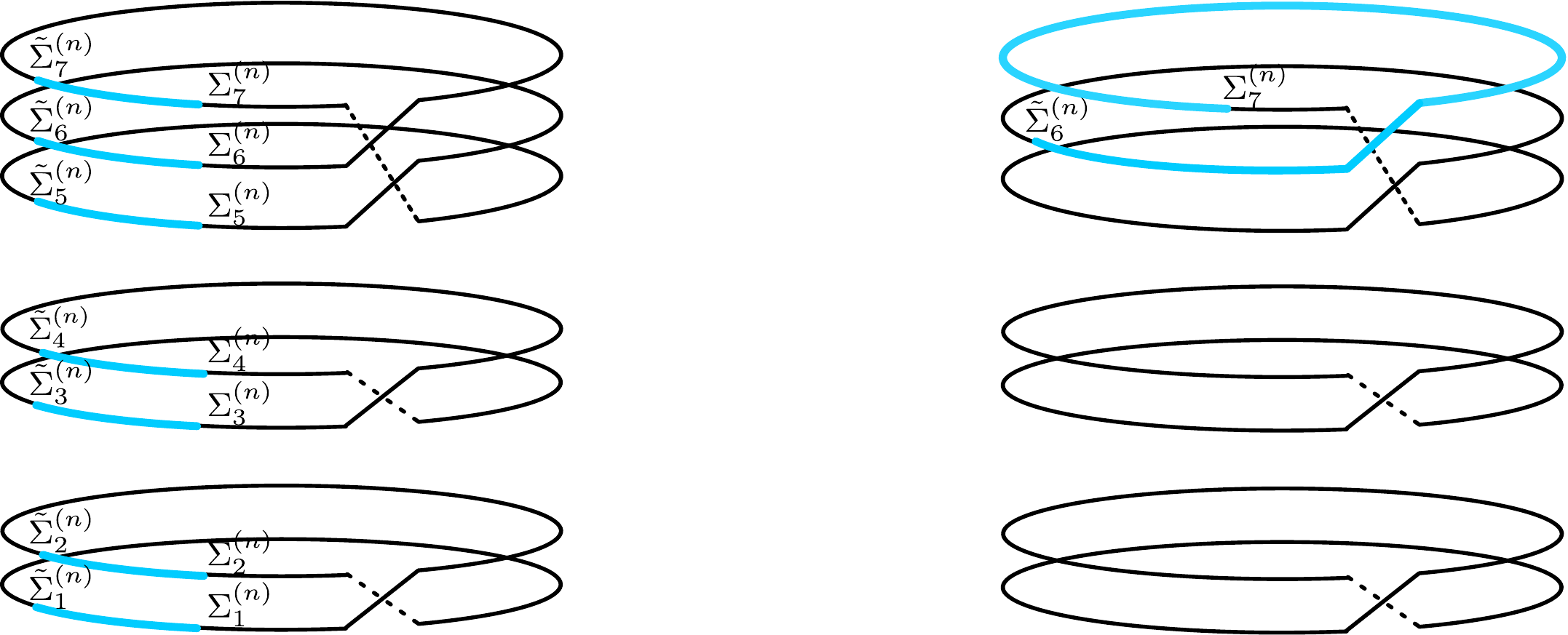}
\caption{\small Multiwound strings, each consisting of  $m$ strands. There are $N_m$  $m$-wound strings such that the total number of strands is $N$. Here we depict $N=7$, $N_2=2$, $N_3=1$.  {\it (Left)} The entanglement entropy is computed by inserting R\'enyi twist operators at the endpoints of the interval on every strand. The entangling region can be visualized in the long string picture as a union of disjoint intervals on all strands. {\it (Right)} Configuration of twists corresponding to the bilocal operator of single interval entwinement. The entangling region extends across different strands of the 3-cycle.} \label{fig:mwound}
\end{figure}

Now take the CFT to be a symmetric orbifold with target space $M^N/S_N$ for some $M$.    As discussed above, this CFT can be regarded as having $N$ elementary strands spliced together into series of cycles (``long strings") determined by the twisted sector.   The twist operator can therefore be regarded as a product of elementary twists 
\begin{equation}
\Sigma^{(n)} = \Sigma^{(n)}_1 \Sigma^{(n)}_2 \cdots \Sigma^{(n)}_N,
\label{entangtwist}
\end{equation}
where each $\Sigma^{(n)}_i$ splices together the $n$ replica copies of the $i$th strand. 
Each elementary twist is in the fundamental representation of $S_N$.
Thus we can write for any $g \in S_N$
\begin{equation}
g[\Sigma^{(n)}_i] \equiv g  \Sigma^{(n)}_i g^{-1} =  \Sigma^{(n)}_{g(i)},
\end{equation}
where $g(i)$ is the strand produced by permuting $i$ by the action of $g$.   The twist operator appearing in the computation of entanglement entropy (\ref{entangtwist}) is a product of all the elementary twists and hence is invariant under the action of $S_N$. The twists configuration that computes the entanglement entropy  is depicted in figure~\ref{fig:mwound} (left). 

We can define entwinement formally in terms of the elementary twists.   
Take $\Sigma_i$ to be an elementary twist operator for strand $i$  and consider the bilocal combination 
\begin{equation}
\tilde\Sigma^{(n)}_i\!(1) \, \, {\Sigma}^{(n)}_i\!(e^{i(\alpha + 2\pi \ell)} )
\end{equation}
where for compactness we have only written the holomorphic coordinate and $\tilde\Sigma_i$ is the conjugate twist.   In this bilocal operator the $\Sigma_i$ is taken around the complex plane relative to $\tilde \Sigma_i$  by an amount $\alpha + 2\pi \ell$.  In a specific state of the symmetric orbifold the $i$th strand is generally spliced with $k$ other strands into a long string.   In such a state the twist $\Sigma$ will be inserted on a different strand, as represented on the right of figure~\ref{fig:mwound}. We then consider the bilocal, gauge-invariant quantity
\begin{equation}
\frac{1}{|S_N|} \sum_{g \in S_N} \tilde\Sigma^{(n)}_{g(i)}\!(1) \, \, {\Sigma}^{(n)}_{g(i)}\!(e^{i(\alpha + 2\pi \ell)} )\,,
\label{Tdef}
\end{equation}
where $|S_N|$ is the cardinality of $S_N$, $i$ is any reference strand of the CFT, and $g(i)$ is the strand to which $i$ is transported when all the strands are permuted by $g \in S_N$.  Because we are summing over all permutations in $S_N$, the final quantity is independent of $i$.   Its expectation value computes the R\'enyi analog of entwinement of single intervals. 

When $\ell>k$  ($k$ being the number of strands of the specific cycle, i.e.\ long string, in which the strand $g(i)$ lives)  we mean the operator in  the sum in (\ref{Tdef}) to represent the twisted boundary conditions of the replicated set of fields on the full long string. Intuitively, we can imagine starting from a short interval on a single strand and putting twisted boundary conditions on the fields inside the interval. Enlarging the interval until it eventually covers the full string, i.e.\ $\alpha+2\pi \ell=2\pi k$, represents putting twisted boundary conditions on the fields on the full string. Further increasing the interval to $\alpha+2\pi \ell>2\pi k$ does not change this picture and just keeps all fields on the long string twisted, nothing more. Keeping this in mind, we can define the entwinement as
\begin{equation}\label{entwinement_single}
E_\ell(\alpha) = \lim_{n\to 1} \frac{1}{1 - n} \log \left[ \bra{\Psi}  \frac{1}{|S_N|} \sum_{g \in S_N} \tilde\Sigma^{(n)}_{g(i)}\!(1) \, \, {\Sigma}^{(n)}_{g(i)}\!(e^{i(\alpha + 2\pi \ell)} ) \ket{\Psi} \right].
\end{equation}
For symmetric product orbifolds this is just a formal way of saying that we are calculating the entanglement entropy of a connected set of partial strands in a long string.   This definition of single interval entwinement can be generalized to multi-interval entwinements by taking a product of operators like (\ref{Tdef}) defined at different locations and strands.  A particular example of multi-interval entwinement is entanglement, where we take a product of the same interval  with $\alpha < 2\pi$ in each of the $N$ strands.

The formalism described above  is  general.   We can consider theories with any discrete gauge symmetry $H$, and elementary twist operators in any representation $R$ of $H$ that is useful.    These can be used to define bilocal, gauge invariant twist operators of the form (\ref{Tdef}) and products of such operators.    Entwinements defined as expectation values of these quantities are a very general new class of gauge-invariant objects than can be used to characterize quantum wavefunctions in two-dimensional theories.   
Conceptually we can also talk about the entanglement of subsets of degrees of freedom in a spatial region even in higher dimensional theories on any manifold, but we need a formalism for calculating such quantities efficiently.

In the next section we will use the replica method to explicitly compute entwinements in symmetric orbifold conformal field theories in two dimensions.

\setcounter{equation}{0}
%%%%%%%%%%%%%%%%%%%%%%%%%%%%%%%%%%%%%%%%%%%%%%%%%%%%%%%%%
\section{Entwinement in symmetric orbifold CFTs}
\label{sec:orbifold}

In the following we analyze symmetric orbifold CFTs on a circle of length $L$. These are obtained starting with a seed CFT with target space $M$ and central charge $c$. 
The orbifold theory  $\mC =$ CFT$^{N}/S_N$ has target space $M^{N}/S_N$ and central charge $c_N=cN$. Because of the $S_N$ identification, states need only be periodic up to the action of a group element.  In a sector twisted by $h \in S_N$, the boundary conditions are
\be
\phi_{i}(L)=  \phi_{h(i)}(0) ~ \qquad {i=1,\dots ,N} \, ,
\ee
where here $\phi_i$ collectively indicates the fields in the $i$-th copy of the CFT. All physical states should be invariant under the action of $S_N$.  Since acting with a group element $g$ maps the sector twisted by $h$ to that twisted by $ghg^{-1}$, twisted sectors should really be labeled by conjugacy classes $[h]$, as mentioned in the previous section.

Twisted states can be conveniently  obtained through the action of {\it orbifold twist operators} on untwisted states.  An orbifold twist operator $\sigma_{m}(0)$  at the origin of the complex plane 
causes $m$ copies of the target space $M$ to be linked together by the periodicity condition
\be
\sigma_{m}(0): \qquad\phi_j( z  e^{2 \pi i })  = \phi_{j'}( z ) ~;~~~j' = (j+1)\mod m\, .
\ee
Here  $j = 1,\ldots,m$, and the twist operator can be thought of as linking $m$ strands of string, each with period $2\pi$, into a single long string with period $2m\pi$.
We will be interested in twisted states of the form 
\be \label{eq:state}
|\psi \rangle = \prod_{m=1}^{N} \LT\sigma_{m} (0) \RT^{N_m} |0 \rangle\, , 
\ee
where each $\sigma_m$  acts on a different subset of the $N$ copies of the target space $M$.   Thus, there will be $N_m$ long strings of period $2m\pi$ and $\sum_m m N_m= N$. In radial quantization this prepares a state on the spatial circle, and  the corresponding out state is
 \be 
\langle \psi | = \langle0|  \prod_{m=1}^{N} \left[ \tilde\sigma_{m} (\infty) \right]^{N_m}\equiv \langle0|  \prod_{m=1}^{N} \lim_{z, \bz \to \infty } \left[  z^{2 h_m} \bz^{2 \bar h_m} \tilde\sigma_{m} (z, \bz) \right]^{N_m} \, , 
\ee
where $\tilde\sigma_m$ has opposite action to $\sigma_m$. The twists transform as primaries with conformal weights 
\be
h_m = \bar h_m=\frac{c}{24}\(m -\frac{1}{m}\) \ .
\ee
We wish to define entwinement in such twisted states of the symmetric product CFT.

In radial quantization we can specify  a connected entangling region at a fixed time $t=0$ by an arc on the unit circle.    
As explained in the previous section, entwinement in twisted states of the form \eqref{eq:state} can be computed using the replica trick, by inserting \emph{elementary replica twist operators}, $\Sigma_i$. Such twists act each on a single strand out of the $N$ strands in the CFT and can be thought of as connecting  the $n$-fold R\'{e}nyi replicates of that strand.  
The elementary replica twists, $\Sigma_i$, have conformal weights 
\be
H_n =\bar H_n= \frac{c}{24}\(n-\frac{1}{n}\) \, . 
\ee
We  insert the elementary twists at the endpoints of the chosen interval of length $\alpha+2\pi\ell$  and average over the symmetric group, which moves the left boundary of the interval over all strands while keeping the length of the interval fixed. The  state of the replicated theory, $|\Psi\rangle$,  is obtained inserting orbifold twist operators  for each of the $n$ replica copies of the theory,
\be
\ket{\Psi}=
 \[ \prod_{m=1}^{N}~ [  \sigma_{m} (0) ]^{N_m} \]^n |0 \rangle \, ,
\ee
i.e., by taking products of (\ref{eq:state}) for each of the $n$ replicated theories.
 In terms of $|\Psi\rangle$, we can write
 \be \label{eq:entw}
E_{\ell}(\alpha)=\lim\limits_{n\rightarrow 1}\frac{1}{1-n} \log \[  \frac{1}{| S_N |}\sum_{g\in S_N} \langle \Psi | \tilde \Sigma^{(n)}_{g(i)}(1) \Sigma^{(n)}_{g(i)}(e^{2\pi i \ell}x,e^{-2\pi i \ell}\bx) | \Psi  \rangle \]\, , 
\ee
where we have taken the entwinement interval to extend between $1$ and $x $ in the complex plane; in the notation of the previous section, $x=e^{i\alpha}$.  As before, since we are averaging over $S_N$, the result is independent of the arbitrary choice of the initial strand $i$, which can run from $1$ to $N$.
This correlator is a four-point function in the cyclic orbifold theory $\mC^n/\Z_n$ obtained taking $n$ replicas of the original symmetric orbifold  $\mC=$ CFT$^{N} /S_N$. 

At the practical level then,  the computation above simplifies drastically.  Consider a term in (\ref{eq:entw}) where the strand $g(i) = k$ is embedded in a particular long string with $m$ strands, which we relabel here as $1,2,\ldots, m$ for convenience.   Let us define a notation
\begin{equation}
\Sigma_k^{(n)}(e^{2\pi i \ell} \, x , e^{-2\pi i \ell} \bar{x}) \equiv
\Sigma_{k+\ell}(x,\bar{x}),
\end{equation}
where the subscripts $k$ and $k+\ell$ are understood modulo $m$ because of the cyclic symmetry of the $m$-stranded long string.
Then we  have to  compute terms like
\be \label{eq:scyclcntr}
\langle \Psi | \tilde \Sigma^{(n)}_{k}(1)  \Sigma^{(n)}_{k+\ell}(x,\bx) | \Psi \rangle  = \langle \[ \tilde\sigma_m(\infty) \]^n \tilde \Sigma^{(n)}_{k}(1) \Sigma^{(n)}_{k+\ell}(x,\bx) \[ \sigma_m(0) \]^n  \rangle \, .
\ee
As discussed in the previous section, if the long string is shorter than the interval in question, i.e., $2\pi m<\alpha+2\pi\ell$, (\ref{eq:scyclcntr}) is understood as computing 
 how the entire long string is entangled with the rest of the system.
 
%%%%%%%%%%%%%%%%%%%%%%%%%%%%%%%%%%%
\subsection{Single interval entwinement}
\label{sec:Einterval}

\begin{figure}[t]
\centering 
\includegraphics[width=0.7\textwidth]{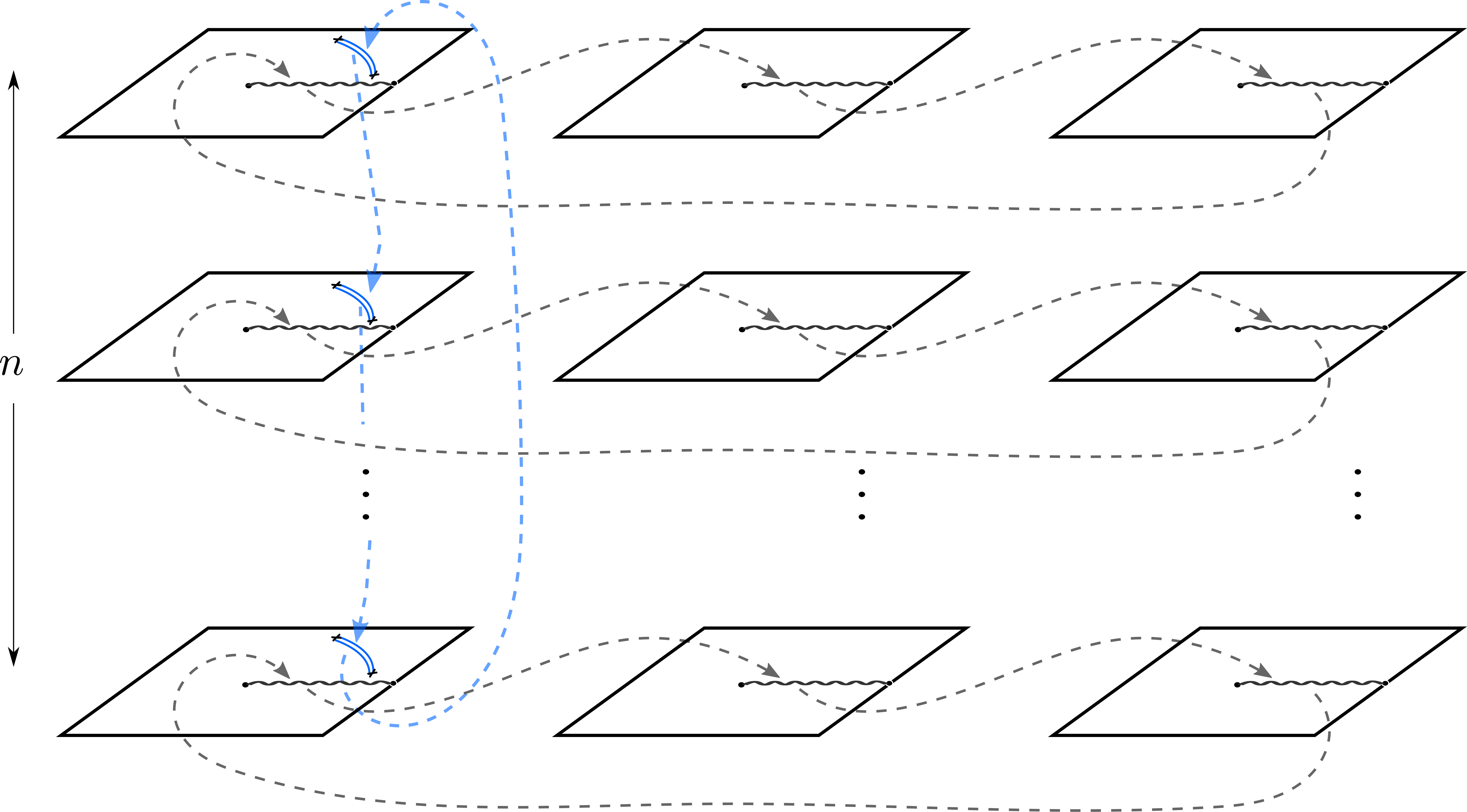}\\
 \hfill \caption{\small   Representation of the branching structure of a correlator of the form  \eqref{eq:correlator} in the simple case of a single strand entwinement, $\ell=0$, on a $3$-cycle factor, $m=3$.} \label{fig:cover}
\end{figure}

To evaluate the entwinement of a single interval we must therefore compute the correlator
\be \label{eq:correlator}
\langle [ \tilde  \sigma_m (\infty)]^n \tilde \Sigma^{(n)}_{k}(1)  \Sigma^{(n)}_{k+\ell}(x,\bar x) [\sigma_m (0)]^n  \rangle \ .
\ee
The branching structure of the correlator is effectively $m n$-dimensional as illustrated  in figure~\ref{fig:cover}, and we can label the twist fields in terms of these $m n$ sheets involved in the correlator:
\bea
\LT\sigma_{m}\RT^n &=& \sigma_{\(1\ldots m\)}\sigma_{\(m+1\ldots 2m\)}\ldots\sigma_{\(m(n-1)+1\ldots mn\)}\,, \\
\Sigma^{(n)}_{k} &=& \sigma_{\( k,k+m,\ldots, k+(n-1)m\) } 
\eea
in the cycles notation $\sigma_{(\ldots)}$ of $S_{m n}$. 

The correlator \eqref{eq:correlator} can  be evaluated  through a uniformization  map to a covering space. To compute the map we extend a calculation of \cite{Asplund:2011cq}, which dealt with the case $m=2$.   We first observe that the branched covering has spherical genus, as determined by the Riemann-Hurwitz formula
\be
g = \frac{1}{2}\sum\limits_{i} (r_i- 1) - s+1 = \frac 1 2 \left[ 2 n (m-1) + 2(n-1)\right] - m n +1 = 0\, .
\ee
The sum is over the twist insertions in \eqref{eq:correlator} of length $r_i$, and  $s = mn$ is the total number of sheets involved in the correlator. An explicit formula for such twist correlation function in terms of the properties of the uniformization map to a genus zero surface is worked out in appendix D of \cite{Avery:2010qw} (see also \cite{Lunin:2000yv,Lunin:2001pw}). 
In appendix~\ref{app:twist}, we review these results and apply them to the computation of single interval entwinement. 
Our final result is (see eq.~\eqref{eq:resultcorrelator} in the appendix)
\begin{align} \label{eq:resultcor}
\langle [ \tilde  \sigma_m & (\infty)]^n  \tilde \Sigma^{(n)}_{k}(1)  \Sigma^{(n)}_{k+\ell}(x,\bar x) [\sigma_m (0)]^n  \rangle  = 
\left[ m^2 \left| A\right|^{m-1} \left|A-1\right|^2\right]^{-\frac{c}{12}\LF n-\frac{1}{n}\RF}\,,
\end{align}
with $A= x^{1/m}e^{2\pi i\ell/m}$ and $c$ the central charge of a single copy. 

To obtain the entwinement of an interval of opening angle $\alpha$ on the spatial circle $w \sim w+ L$, we only need to relate the result    \eqref{eq:resultcor} obtained on the plane  to the computation on the cylinder, with elementary replica twists inserted at $w=0$ and $w=\alpha L/(2\pi)$.  Using the map $z = e^{ \frac{2\pi i w}{L}}$:
\be
\langle \Psi| \tilde \Sigma^{(n)}_{k}(0)  \Sigma^{(n)}_{k+\ell} \Big(\frac{\alpha L}{2\pi}\Big) | \Psi  \rangle_{\rm cylinder} = \(\frac{2 \pi}{L}\)^{4 H_n} \langle \Psi|  \tilde \Sigma^{(n)}_{k}(1) \Sigma^{(n)}_{k+\ell}(x,\bx) | \Psi  \rangle 
\ee 
with $x= e^{i \alpha} $, $\bx =e^{- i \alpha}$.  If we explicitly insert a UV cutoff $\epsilon_{UV}$ to regulate the twist operators and work out the sum over all elements in $S_N$, we obtain
\be 
E_{\ell}( \alpha) =\lim\limits_{n\rightarrow 1}\frac{1}{1-n}\log \left[ \frac{1}{N}\sum\limits_{m=\ell+1}^{N} m N_m  \left|\frac{m L}{ \pi \epsilon_{UV}}\sin\LF \frac{\alpha+2\pi \ell }{2m}\RF\right|^{-\frac{c}{6}\LF n-\frac{1}{n}\RF}\RT  + O(\epsilon_{UV}^0)\, .  
\label{eq:EWcyl}
\ee
The sum in the first term extends over long strings with $\ell + 1$ and more strands, because shorter strings are completely covered by intervals of length $\alpha + 2\pi \ell$.   The contribution from each of these shorter strings computes the entanglement of the string with the rest of the theory.   Since the short string is by construction disconnected from all other strands in this particular twisted sector, its entanglement entropy will not have the dominant UV-divergent contributions that are present for the longer strings with $\ell + 1$ and more strands.
In formulas, if one sets $\alpha+2\pi\ell\approx 2\pi m$ up to a contribution of order $\epsilon_{UV}$, the occurences of the cutoff will cancel in (\ref{eq:EWcyl}) between $1/\epsilon_{UV}$ and the sine, yielding a cutoff-independent result, which is small compared to the cutoff-dependent terms arising from long strings longer than the interval.

%%%%%%%%%%%%%%%%%%%%%%%%%
\subsection{Entanglement entropy of a spatial region}

The entanglement entropy of an interval 
is a specific case of computing entwinement. 
 The gauge invariant twist operators are decomposed into products of twist operators on each strand as in \eqref{entangtwist}, and the entanglement entropy of a spatial interval can be expressed as
\be 
S(\alpha) = \lim\limits_{n\rightarrow 1} \frac{1}{1-n} \log \LT \bra{\Psi} \tilde \Sigma_1^{(n)}(1) {\Sigma}_1^{(n)}(x,\bar{x}) \tilde\Sigma_2^{(n)}(1){\Sigma}_2^{(n)}(x,\bar{x})\ldots\tilde\Sigma_N^{(n)}(1) {\Sigma}_N^{(n)}(x,\bar{x})\ket{\Psi}\RT.
\ee
In the perspective of entwinement, entanglement entropy coincides with the entwinement of the union over all strands of an interval that fits within a single strand.
Clearly, as this quantity is already gauge invariant, the sum over $S_N$ appearing in the entwinement definition exactly cancels the normalization $|S_N|^{-1}$.

In fact the entanglement entropy for a state of the form \eqref{eq:state} is not known in general. 
For instance the branching structure of the correlator leads generically to a covering space of non-trivial genus, and thus one cannot straightforwardly apply the same techniques we used for computing entwinement.   However, in the limit of a short  interval ($x,\bx \rightarrow 1$), using the OPE of the elementary twists $\tilde \Sigma^{(n)}_i(1){\Sigma}^{(n)}_i(x,\bar{x})\sim \mathbbm{1}/ |1-x|^{2H_n}$ the correlator factorizes. Via the conformal map to the cylinder,  the result reproduces the short interval expansion of the entanglement entropy for a CFT on a circle of length $L$ with central charge $c_N =c N$,
\be
S(\alpha) \approx\frac{ c_N }{3} \log \frac{\alpha L}{\pi\epsilon_{UV}} \, .
\ee
This is $N$ times the single strand ($\ell=0$) short interval expansion ($\alpha \to 0$) of the entwinement result given in \eqref{eq:EWcyl}. 
This reflects the fact that $\ell=0$ entwinement computes the entanglement for a single factor in the symmetric product orbifold theory, while entanglement entropy simultaneously involves fields in all $N$ factors.

%%%%%%%%%%%%%%%%%%%%%%%%%%%%%%%%%%%%%%%%%%%%%%%%%%%%%%%%%
%%%%%%%%%%%%%%%%%%%%%%%%%%%%%%%%%%%%%%%%%%%%%%%%%%%%%%%%%

\setcounter{equation}{0}
\section{D1-D5 CFT}
\label{sec:D1D5}

A well-known example of a symmetric orbifold CFT is the D1-D5 CFT. This is realized in type IIB string theory compactified on $S^1 \times T^4$ (or $S^1 \times K3$), with $N_1$ D1-branes wrapping the circle and $N_5$ D5-branes wrapping the entire compact product space. 
The near horizon geometry of the D1-D5 brane system is AdS$_3 \times S^3 \times T^4$, and one can formulate a two-dimensional CFT at the conformal boundary of the AdS$_3$. This is an ${\cal N}=(4,4)$ supersymmetric sigma model with $SU(2) \times SU(2)$ R-symmetry, corresponding to the isometry group of the $S^3$, another $SU(2) \times SU(2)$ global symmetry and central charge equal to $ 6 N_1 N_5$ (see for instance \cite{Avery:2010qw} for a review). The moduli space of the CFT contains an orbifold point where the theory consists of  $N \equiv N_1N_5$ copies of a $c = 6$ free CFT of 4 real bosons and their fermionic superpartners with target space $(T^4)^N / S_N$. 

We will work at the orbifold point of the D1-D5 CFT and focus on the Ramond ground states. These can be constructed by multiplying together bosonic and fermionic twist operators to achieve a total twist of $N$. The theory contains eight bosonic and eight fermionic twists labeled in terms of the global symmetries. Since we are only interested in computing correlators of bosonic quantities that do not carry R-charge, we can simplify the discussion and generically consider the normalized symmetric orbifold microstates \eqref{eq:state}. In this section we will consider two examples of such states, which in the large $N$ limit are dual to conical defects and zero mass BTZ black holes in the bulk (see \cite{Balasubramanian:2000rt} for  discussion of the map between Ramond ground states of the D1-D5 system and AdS$_3$ gravity).\footnote{The BTZ black hole is actually dual to an ensemble of states; we will comment on this point and on the notion of typical states in section \ref{sec:BTZ}.} Points in the moduli space with a geometric supergravity description are actually far from the orbifold point where we perform our computations,
and agreement with semiclassical gravity is not to be expected a priori. However protected BPS quantities can be computed exactly at the orbifold point and it has been proposed that agreement should extend also to observables computed in terms of covering space constructions \cite{Martinec:2002xq}.

%%%%%%%%%%%%%%%%%%%%%%%%%%%%%%%%%%%%%%%%%%%%%%%%%%%%%%%%%

\subsection{Conical defects}

Simple Ramond ground states of the D1-D5 CFT are of the form \cite{Lunin:2000yv, Lunin:2001pw, Balasubramanian:0508}
\be
\ket{\psi} = \left[ \sigma_{\tilde m}(0)\right]^{N/ \tilde m} \ket{0}\,,
\label{ConicalDefectState}
\ee
where $N/\tilde m$ is an integer. 
The holographically dual bulk geometries are conical defects
\be \label{eq:conicadefectmetric}
ds^2 = -\( \frac{1}{\tilde m^2} + \frac{r^2}{R_{\rm AdS}^2}\) dt^2 + \( \frac{1}{\tilde m^2} + \frac{r^2}{R_{\rm AdS}^2}\)^{-1} dr^2 + r^2 d\theta^2,
\ee
where $\theta$ is a periodic coordinate and the AdS radius is related to the length of the spatial circle introduced in the previous section via $R_{\rm AdS} = L/(2\pi)$. These geometries can be obtained from empty AdS$_3$ in global coordinates via $\mathbbm{Z}_{\tilde m}$ identifications. 
For a single interval of opening angle $\alpha$ on the boundary, there exist multiple geodesics connecting the endpoints of the interval. These have regulated lengths \cite{Balasubramanian:2014sra}
\be
{\cal L}_\ell (\alpha)=2 R_{\rm AdS} \log \LT \frac{2 \tilde m r_\infty}{R_{\rm AdS}} \sin\LF \frac{\alpha+2\pi \ell}{2 \tilde m}\RF\RT \, .
\ee
The index $\ell=0,\ldots,\tilde m-1$ counts the number of times the geodesics winds around the conical defect at $r=0$.
Here $r_\infty$ is an IR gravitational cutoff. Identifying it with the field theory cutoff $\epsilon_{UV}$ via $r_\infty = \frac{L}{2 \pi} \frac{R_{\rm AdS}}{\epsilon_{UV}}$ and substituting $c=6$ in \eqref{eq:EWcyl}, we find the relation between entwinement in the state \eqref{ConicalDefectState} and geodesic lengths
\be
E_{\ell}(\alpha) =\frac{{\cal L}_\ell (\alpha)}{R_{\text{AdS}}}.
\label{EntwineVersusGeodesic}
\ee
Therefore in this case, the field theory notion of entwinement computes the length of  geodesics in AdS, in agreement with the idea advanced in~\cite{Balasubramanian:2014sra} that non-minimal geodesics in the bulk capture the entanglement of internal degrees of freedom. Ordinary entanglement entropy on the other hand is related to the length of minimal geodesics, in agreement with the Ryu-Takayanagi formula~\cite{Ryu:2006bv}
\be
S (\alpha)= \frac{{\cal L}_0 (\alpha)}{4 G_N}.
\ee
Since the D1-D5 central charge is $c N = 3 R_{\text{AdS}}/(2 G_N)$, the entanglement entropy of an interval coincides with $N$ times the $\ell = 0$ single interval entwinement $E_{0}$. 

%%%%%%%%%%%%%%%%%%%%%%%%%%%%%%%%%%%%%%%%%%%%%%%%%%%%%%%%%

\subsection{Zero mass BTZ black hole}\label{sec:BTZ}

One could wonder whether the length of long geodesics also captures entwinement in a black hole background. 
We will show this is indeed the case in the zero mass BTZ black hole
\be
ds^2 = - \frac{r^2}{R_{\text{AdS}}^2} dt^2 + \frac{R_{\text{AdS}}^2}{r^2} dr^2 + r^2 d\theta^2\,,
\ee
which arises in the $\tilde m \to \infty$ limit of the conical defect geometry \eqref{eq:conicadefectmetric}.   The $M=0$ black hole has a horizon of zero size which coincides with the singularity.

The $M=0$ black hole geometry is not dual to a particular CFT microstate, but rather to an ensemble of states of the D1-D5 CFT with fixed $N$.
Following \cite{Balasubramanian:0508}, instead of working with the microcanonical ensemble, it is more convenient to work in the canonical description where $N_m$ as well as the total number of strands fluctuate, but the ensemble average is fixed to $N$.   The average number of $m$-cycles is \cite{Balasubramanian:0508}
\begin{equation}
\langle N_{m} \rangle = \frac{8}{\sinh\beta m}\,,
\label{eq:TypicalState}
\end{equation}
where the inverse fictitious temperature $\beta$ is determined in terms of the average $N$ of the total number of strands, as
\begin{equation}
N = \Bigg \langle\sum_{m=1}^{\infty} m N_m \Bigg\rangle \simeq \frac{2\pi^2}{\beta^2}\,. 
\end{equation}
In the large $N$ limit, typical states in the ensemble have individual twist distributions that lie very close to \eqref{eq:TypicalState} and expectation values of observables in a typical state deviate by only a small amount from those computed in the ensemble. In the following, we therefore compute entwinement in a typical state with representative distribution \eqref{eq:TypicalState}, rather than in the ensemble.  

The single interval entwinement in a typical microstate is again given by~(\ref{eq:EWcyl}).
As we prove in appendix~\ref{SqrtN}, for fixed $\alpha$ and $\ell$, the sum over $m$ is dominated by the terms with $m \sim O(\sqrt{N})$. For $\ell \ll \sqrt{N}$, we obtain
\be
E_{\ell}(\alpha)\approx \frac{c}{3} \log \LT\frac{L}{2\pi\epsilon_{UV}}(\alpha+2\pi \ell)\RT.
\ee

Black holes admit a region outside the horizon which is not penetrated by minimal geodesics, which we can call the entanglement shadow of the black hole. Just as in the case of the conical defect, non-minimal geodesics penetrate the entanglement shadow. The non-minimal geodesics wind around the horizon and the bigger their winding number, the closer they wrap the horizon. The $M=0$ BTZ black hole has a horizon that shrinks to zero size, but nevertheless it has a finite entanglement shadow. The lengths of non-minimal geodesics in the $M=0$ BTZ black hole background are given by
\be
{\cal L}_\ell(\alpha)= 2R_{\rm AdS}\log\LT\frac{r_\infty}{R_{\rm AdS}}(\alpha+2\pi \ell)\RT.
\ee
Using again that $c=6$ and $r_\infty = \frac{L}{2\pi}\frac{R_{\rm AdS}}{ \epsilon_{UV}}$, we recover that entwinement and geodesic length are related by 
\be 
E_\ell(\alpha) = \frac{{\cal L}_\ell(\alpha)}{R_{\rm AdS}}.
\ee
As $\ell \to \sqrt{N}$, the relation between  entwinement and geodesic length breaks down.  But the corresponding long geodesics, which wind very many times around the black hole, approach the horizon to within a Planck length.   It is not clear that they are well defined in the quantum theory where we expect classical geometry to be ill-defined at the Planck scale.  Hence it is not surprising that the lengths of these geodesics  do not match the corresponding entwinements.

%%%%%%%%%%%%%%%%%%%%%%%%%%%%%%%%%%%%%%%%%%%%%%%%%%%%%%%%%%%%%%%%%%%%%%%%%

\setcounter{equation}{0}
\section{Discussion and outlook}
\label{conclusions}

We have verified the correspondence between single-interval entwinement and lengths of non-minimal geodesics in conical defects and $M=0$ BTZ black holes, which correspond to specific states of the D1-D5 system.   We studied these configurations because there are explicit constructions of the corresponding states in the literature \cite{Balasubramanian:0508}.  Furthermore, these states are BPS-protected ground states in the Ramond sector of the theory \cite{Balasubramanian:2000rt,  Martinec:2002xq}, so that we can expect non-renormalization of some quantities as we deform the theory away from its orbifold point.   Of course this does not mean that all correlation functions extrapolate from weak to strong coupling, but we were essentially computing partition sums after a conformal map, which might help explain the agreements we found between field theory and gravity.  It  has also been seen that certain graviton correlators computed at the orbifold point do match the gravitational results which are related to the strongly coupled theory \cite{Balasubramanian:0508}.   Possible general reasons for such matching are discussed in \cite{Martinec:2002xq}.  It should be possible to extend our computations in at least two interesting directions.  First, we can consider the general R-charged Ramond ground states of the D1-D5 CFT.  The $M=0$ BTZ black hole and the conical defects considered here were two specific examples of such states, but a more general class is discussed in \cite{Balasubramanian:0508}, including candidate states describing ``black ring'' geometries.   Another interesting extension is to consider rotating, but extremal, AdS$_3$ black holes.   These are represented in the D1-D5 CFT by adding energy to the left moving sector alone.  One approach would be to add a small left-moving temperature; another would be to perturb the theory with a holomorphic stress tensor.  Both of these are settings where it would be very interesting to compute both spatial entanglement and entwinement.

While in this paper we have mostly focused on 2d symmetric orbifold CFTs, our definition \eqref{entwinement_single} of entwinement can be extended to more general discretely gauged theories. Consider, for instance, the discrete rotation orbifold $\Rbar^2/\Zbar_N$ with $N>2$. In the sector twisted by the generator of $\Zbar_N$, the target space coordinates satisfy 
\bea
X(\tau,2\pi)&=&\cos\left(\frac{2\pi}{N}\right)X(\tau,0)-\sin\left(\frac{2\pi}{N}\right)Y(\tau,0),\\
Y(\tau,2\pi)&=&\sin\left(\frac{2\pi}{N}\right)X(\tau,0)+  \cos\left(\frac{2\pi}{N}\right)Y(\tau,0).
\eea
We can extend the range of the $\sigma$ coordinate in $X(\tau,\sigma)$ to $0\leq\sigma<4\pi$ by defining
\be
X(\tau,\sigma)\equiv\cos\left(\frac{2\pi}{N}\right)X(\tau,\sigma-2\pi)-\sin\left(\frac{2\pi}{N}\right)Y(\tau,\sigma-2\pi), \ \ \ 2\pi\leq\sigma<4\pi.
\ee
A twisted sector string is then determined by the ``long string'' profile $X(\tau,\sigma)$ with $0\leq\sigma<4\pi$, which satisfies the boundary condition
\be
X(\tau,4\pi)=-X(\tau,0)+2\cos\left(\frac{2\pi}{N}\right)X(\tau,2\pi).
\ee 
Entwinement defined in analogy with \eqref{entwinement_single} then quantifies how one part of this long string is entangled with its complement. There does seem to be an important difference, however, with symmetric orbifold theories. For the above rotation orbifold, we could equally well have defined entwinement by considering target space coordinates $X',Y'$ related to $X,Y$ by rotation in field space over an arbitrary angle. Entwinement defined using long string profiles $X'(\sigma)$ would generically be different from that defined using $X(\sigma)$, reflecting the basis dependence of measures of quantum entanglement in general.   In contrast, for symmetric orbifolds the requirements that the target space coordinates should be mapped into each other by permutations and should have diagonal kinetic terms does select a preferred notion of entwinement. It is  interesting to ask for which gauge theories our definition of entwinement leads to ``natural'' quantities, including quantities with a clear geometrical meaning in a holographic dual. Other generalizations worth studying are continuous gauge theories, higher dimensional theories and matrix models.

Note that entwinement, as we defined it, measured the lengths of geodesics in units of the AdS scale, rather than in units of the Planck length (which is related to the three dimensional Newton constant $G_N$).   In the original formulation of holographic entanglement entropy \cite{Ryu:2006bv,  Ryu:2006ef}, it seemed natural that $G_N$ should appear in the  formulas, in analogy with black hole entropy which is measured by horizon area in units of $G_N$.   Of course field theories with a holographic dual having a classical description generally have a large number of local degrees of freedom arising from e.g.  dynamical variables that are large matrices.   The entwinement that we are defining seeks to piece apart the entanglement of some of these local degrees of freedom (e.g. parts of the local matrices) with other elements of the Hilbert space.  As such, we are extracting the elementary parts out of which spatial entanglement arises in such quantum field theories.  These elementary entanglements are correspondingly smaller, and are thus related to geodesic lengths measured relative to a length scale that is much larger than the Planck length.     One might ask if single interval entwinement can ever be of the same order of magnitude as the spatial entanglement as we make the interval size larger.   At least in the 2d CFT states we considered, this cannot happen because entanglement between different long strings is negligible and within a single long string entanglement only depends logarithmically on the interval size.   This could be different in excited or thermal states where there could be extensive contributions to entanglement, or in theories with less local interactions such as matrix models.   It is also interesting that the fundamental object (\ref{entwinement_single}) from which we construct entwinement is non-local.  This recalls the discussion in \cite{polchinski} of the relevance of non-local observables in field theory for reconstructing local physics in AdS space in a gauge-invariant manner.   

One of our  goals in this paper has been to define the field theoretic dual of extremal, non-minimal geodesics in AdS$_3$.  These geodesics also appear in the semiclassical CFT computation of R\'{e}nyi entropies in terms of the conformal block expansion of heavy-heavy-light-light correlators  \cite{Asplund:2014coa}.  There it was shown that the single interval entanglement entropy in a state created by heavy operator insertions, and dual to an AdS$_3$ conical defect or BTZ, is well approximated by the semiclassical identity block and reproduces the Ryu-Takayanagi minimal geodesic result. This leading answer for the four-point function however  has monodromies as the R\'{e}nyi twists are moved around the heavy operators in the CFT. These monodromies transform the minimal geodesic result into quantities related to the length of non-minimal geodesics. Therefore in this context non-minimal geodesics are also related in the dual CFT to analytic properties of semiclassical Virasoro blocks.

\setcounter{equation}{0}
%%%%%%%%%%%%%%%%%%%%%%%%%%%%%%%%%%%%%%%%%%%%%%%%%%%%%%%%%%%%%%%%%%%%%%%%%

\section*{Acknowledgments}

We thank Alexandre Belin, Netta Engelhardt, Jutho Haegeman, Volkher Scholz, Norbert Schuch,  Karel Van Acoleyen, Henri Verschelde, and Frank Verstraete for very helpful discussions.   We are also particularly grateful to Bartek Czech, Lampros Lamprou, Sam McCandlish, and Jamie Sully for discussing details of their ongoing work on related questions with us.  This work was supported in part by a grant from the Simons Foundation (\#385592, Vijay Balasubramanian) through the It From Qubit Simons Collaboration, by the Belgian Federal Science Policy Office through the Interuniversity Attraction Pole P7/37, by FWO-Vlaanderen through projects G020714N, G044016N and Odysseus grant G.001.12, by the European Research Council grant no. ERC-2013-CoG 616732 HoloQosmos, by COST Action MP1210 The String Theory Universe, and by Vrije Universiteit Brussel through the Strategic Research Program ``High-Energy Physics''. It was performed in part at the Aspen Center for Physics, which is supported by National Science Foundation grant PHY-1066293. Research at Perimeter Institute is supported by the Government of Canada through Industry Canada and by the Province of Ontario through the Ministry of Research \& Innovation.   T.D.J.\ is Aspirant FWO-Vlaanderen. F.G.\ is a Postdoctoral Researcher of FWO-Vlaanderen and acknowledges support from a JuMo grant of KU Leuven.   V.B., A.B., and F.G.\  thank YITP for support during the program ``Quantum Information in String Theory and Many-body Systems'' and conference ``Quantum Matter, Spacetime and Information''.     V.B.\ and T.D.J.\ also thank the Perimeter Institute for hospitality during the It From Qubit workshop and school.      B.C.\ thanks the organizers of the Nordita program ``Black Holes and Emergent Spacetime'' for hospitality while this work was in progress.

\setcounter{equation}{0}

%%%%%%%%%%%%%%%%%%%%%%%%%%%%%%%%%%%%%%%%%%%%%%%%%%%%%%%%%%%%%%%%%%%%%%%%%

\appendix

\section{Twist correlator}\label{app:twist}
In this appendix we briefly review the result described in \cite{Avery:2010qw} for general bare twists correlators with covering space of spherical genus, and apply them explicitly  to the computation of the correlator in Eq.~\eqref{eq:correlator},
\be \label{eq:correlatorapp}
\langle [ \tilde  \sigma_m (\infty)]^n \tilde \Sigma^{(n)}_{1}(1)  \Sigma^{(n)}_{1+\ell}(x,\bar x) [\sigma_m (0)]^n  \rangle\,,
\ee
where, since the strands are indistinguishable, we have set  without loss of generality $k=1$. 

Consider a general correlator of $P+Q$ normalized twists,
\be \label{eq:correlatorPQ}
\langle \sigma_{p_1} (z_1) \sigma_{p_2}(z_2) \dots \sigma_{p_P}(z_P) \sigma_{q_1}(\infty) \sigma_{q_2}(\infty) \dots  \sigma_{q_Q}(\infty)  \rangle ,
\ee
where $p_i, q_j$ denote the lengths of the symmetric group cycles of the corresponding operators, inserted in the finite $z$-plane and at infinity, respectively. We restrict attention to correlators of genus zero,
\be
g = \frac{1}{2}\sum\limits_{i=1}^{P} (p_i- 1)+\frac{1}{2}\sum\limits_{j=1}^{Q} (q_j- 1) - s+1 =  0\,, 
\ee
where, as in the main text, $s$ denotes the total number of copies. 

The map to the covering space $z= z(t)$ should be meromorphic and reproduce the correct monodromies corresponding to the twist insertions
\be
\begin{array}{lll}
z -z_i \approx a_i (t-t_i)^{p_i}  & \qquad \mbox{ for } z \approx z_i\,, \quad t\approx t_i & \qquad  \forall i=1,\ldots, P \\
z \approx b_1 t^q & \qquad \mbox{ for } z \to \infty \,, \quad t \to  \infty &    \\
z \approx b_j  (t-t_j^{\infty})^{- q_j} &  \qquad \mbox{ for } z \to \infty \,, \quad t\approx t^\infty_j & \qquad  \forall j=2,\ldots, Q
\end{array}
\ee
to leading order near each branch point. The correlator \eqref{eq:correlatorPQ} can then be shown to depend only on the coefficients $a_i,b_j$ and parameters $p_i,q_j$ and to be given by \cite{Avery:2010qw} 
\begin{align} \label{eq:generalcorrelator}
\langle \sigma_{p_1}& (z_1) \sigma_{p_2}(z_2) \dots \sigma_{p_P}(z_P) \sigma_{q_1}(\infty) \sigma_{q_2}(\infty) \dots  \sigma_{q_Q}(\infty)  \rangle  = \nonumber\\
&= \LF \prod\limits_{i=1}^{P} p_i^{-\frac{c}{12}\LF p_i + 1\RF}\RF 
\LF \prod\limits_{j=1}^{Q} q_j^{\frac{c}{12}\LF q_j - 1\RF}\RF
 \LF \prod\limits_{i=1}^{P} \left|a_i\right|^{-\frac{c}{12} \frac{p_i - 1}{p_i}}\RF
  \LF \prod\limits_{j=1}^{F} \left|b_j\right|^{-\frac{c}{12} \frac{q_j+ 1}{q_j} }\RF\left|b_1\right|^{\frac{c}{6}} q_1^{\frac{c}{6}}\,,
\end{align}
where 
\be
F = s + Q - \sum_{j=1}^{Q} q_j
\ee
is the number of distinct images of infinity and $c$ denotes the central charge of  the seed CFT. 

Specializing to the correlator \eqref{eq:correlatorapp}, we have 
\be
\begin{array}{lll}
z \approx a_i (t-t_i)^m  & \qquad \mbox{ for } z \approx 0\,, \quad t\approx t_i & \qquad  \forall i=1,\ldots, n \\
z-1 \approx a_{n+1}  t^n & \qquad \mbox{ for } z \approx 1\,, \quad t\approx 0 &   \\
z- x \approx a_{n+2}  (t-1)^n & \qquad \mbox{ for } z \approx x\,, \quad t\approx 1 &    \\
z \approx b_1 t^m & \qquad \mbox{ for } z \to \infty \,, \quad t\to t_1^\infty = \infty &    \\
z \approx b_j  (t-t_j^{\infty})^{-m} &  \qquad \mbox{ for } z \to \infty \,, \quad t\approx t^\infty_j & \qquad  \forall j=2,\ldots, n 
\end{array}
\ee
and a map that satisfies these properties is
\begin{equation}
z = \LT \frac{ A t^n - (t-1)^n}{t^n - (t-1)^n} \RT^m
\end{equation} 
with 
\be
x = A^m\,.
\ee
In order to completely fix the map that computes the correlator \eqref{eq:correlator}, we choose the branch requiring that for two elementary replica twists in \eqref{eq:correlatorapp} acting on the same strand, i.e.~for $\ell =0$, $A \to 1$ as the interval size goes to zero.  We then impose continuity as we increase the interval size and take $x$ $\ell$ times around the unit circle  in the complex plane,  that is
\be
A = e^{\frac{2 \pi i \ell }{m}} x^{\frac 1 m}  \, .
%= e^{\frac{i(\alpha+2\pi \ell)}{m}}\,.
\ee
In the $t$-plane, the insertions map to
\be
t_k = \frac{1}{1- A^{\frac{1}{n}} e^{\frac{2\pi i k}{n}}}\,, \qquad\qquad t^\infty_j = \frac{1}{1- e^{\frac{2\pi i (j-1)}{n}}}\,,
\label{id1}
\ee
and
\begin{eqnarray}
A t^n - (t-1)^n = (A-1) \prod\limits_{i=1}^{n} (t-t_i)\,, & \quad &\ t^n - (t-1)^n = n \prod\limits_{j=2}^{n} (t-t_j^\infty)\,, 
\label{id2}
\end{eqnarray}
so that the conformal map can also be written as
\begin{equation}
z = \frac{ (A-1)^m}{n^m} \frac{ \prod\limits_{i=1}^{n} (t-t_i)^m}{ \prod\limits_{j=2}^{n} (t-t_j^\infty)^m}\,.
\label{ConfMap}
\end{equation}
From this expression we can directly determine the coefficients
\be \label{eq:bj}
b_1 = \LF \frac{ A-1}{n}\RF^m\,, \qquad b_j = \LT \frac{A-1}{n} \frac{ \prod\limits_{k=1}^{n} (t_j^\infty -t_k)}{\prod\limits_{k=2,k\neq j}^{n} (t_j^{\infty} -t_k^\infty)}\RT^m\, .
\ee
To work out the remaining coefficients $a_i$, we consider the first derivative
\begin{equation}
\frac{dz}{dt} = \frac{ m n (A-1)^m}{\LT t^n - (t-1)^n\RT^{m+1}}t^{n-1} (t-1)^{n-1} \prod\limits_{i=1}^{n} (t-t_i)^{m-1} \,.
\label{id3}
\end{equation}
Near $t\approx 0,1,t_i$ the first derivative vanishes. In fact, the first $m-1$ derivatives vanish near $t=t_i$, and the first $n-1$ near $t=0,1$. This implies that in a Taylor expansion near these points we have 
\be
\begin{aligned}
z &= \frac{1}{m!}\left.\frac{d^m z }{dt^m}\right|_{t=t_i} (t-t_i)^m+ \ldots,\\
z &= 1 + \frac{1}{n!} \left.\frac{d^n z}{dt^n}\right|_{t=0} t^n+ \ldots,\\
z &= x+ \frac{1}{n!} \left.\frac{d^n z}{dt^n}\right|_{t=1} (t-1)^n+ \ldots,
\end{aligned}
\ee
from which we read
\be
\begin{aligned}
a_i &= \frac{n (A-1)^m}{\LT t_i^n - (t_i-1)^n\RT^{m+1}} t_i^{n-1} (t_i-1)^{n-1} \prod\limits_{k=1,k\neq i}^{n} (t_i-t_k)^{m-1}\,,\label{eq:ai}\\
a_{n+1} &= m (-1)^{-(n+1)} (A-1)\,,\\
a_{n+2} &= m A^{m-1}(A-1)\,.
\end{aligned}
\ee
Substituting in \eqref{eq:generalcorrelator}, we obtain
\begin{align}
\langle [ \tilde  \sigma_m & (\infty)]^n   \tilde \Sigma^{(n)}_{1}(1) \Sigma^{(n)}_{1+\ell}(x,\bar x) [\sigma_m (0)]^n  \rangle \nonumber \\
&=m ^{\frac{c}{6} (1-n)} n^{-\frac{c}{6}(n+1) } 
\LF \prod\limits_{i=1}^{n} \left|a_i\right|^{-\frac{c}{12}  \frac{m - 1}{m} }\RF \left|a_{n+1}\right|^{-\frac{c}{12} \frac{n - 1}{n} }
\left|a_{n+2}\right|^{-\frac{c}{12} \frac{n - 1}{n}} \LF \prod\limits_{j=1}^{n}\left|b_j\right|^{-\frac{c}{12} \frac{m + 1}{m} }\RF\left|b_1\right|^{\frac{c}{6}}\,.
\end{align}
To evaluate explicitly the products, observe that from  \eqref{id1}, \eqref{id2}, \eqref{id3} we can derive the following identities
\be
\prod\limits_{k=1}^{n} \left| t_k^n - (t_k-1)^n\right| = \prod\limits_{k=1}^{n} \frac{\left| 1-A\right|}{\left| 1- A^{\frac{1}{n}}e^{\frac{2\pi ik}{n}}\right|^n} \\
= \left|1-A\right|^n\prod\limits_{k=1}^{n} \left|t_k\right|^n 
%&=  \frac{\left|1-A\right|^n}{\left|1-A\right|^n},\\
=1 \,, 
\ee 
\be
\prod\limits_{k=1}^{n} \left| t_k -1\right| 
%&= \prod\limits_{k=1}^{n} \left| \frac{A^{\frac{1}{n}}e^{\frac{2\pi ik}{n}}}{1-A^{\frac{1}{n}}e^{\frac{2\pi ik}{n}}}\right|,\\
= \left|A\right|\prod\limits_{k=1}^{n}\left| t_k\right| 
= \frac{ \left| A\right|}{\left|A-1\right|},\label{id5}
\ee 
\be
\begin{aligned}
\prod\limits_{k=1}^{n}\prod\limits_{j=1,j\neq k}^{n} \left| t_k-t_j\right| &= 
\prod\limits_{k=1}^{n}\frac{1}{\left|A-1\right|} \left| \frac{d}{dt} \LF A t^n -(t-1)^n\RF\right|_{t=t_k}\\
& = \prod\limits_{k=1}^{n} \frac{n}{\left|A-1\right|} \left| A t_k^{n-1} - (t_k-1)^{n-1}\right| \\
%&=\prod\limits_{k=1}^{n} \frac{n}{\left|A-1\right|} 
%\left| \frac{A\LF 1-A^{-\frac{1}{n}}e^{\frac{-2\pi i k}{n}}\RF}{\LF 1- A^{\frac{1}{n}} e^{\frac{2\pi ik}{n}}\RF^{n-1}}\right|,\\
%&=\prod\limits_{k=1}^{n} \frac{n}{\left|A-1\right|} 
%\left| \frac{A^{1-\frac{1}{n}}e^{\frac{-2\pi i k}{n}}}{\LF 1- A^{\frac{1}{n}} e^{\frac{2\pi ik}{n}}\RF^{n-2}}\right|,\\
&=\prod\limits_{k=1}^{n} \frac{n}{\left|A-1\right|} 
\left| \frac{A^{1-\frac{1}{n}}}{\LF 1- A^{\frac{1}{n}} e^{\frac{2\pi ik}{n}}\RF^{n-2}}\right| \\
&=\frac{n^n \left|A\right|^{n-1}}{\left|A-1\right|^n} \prod\limits_{k=1}^{n} \left|t_k\right|^{n-2} 
% &=\frac{n^n \left|A\right|^{n-1}}{\left|A-1\right|^n} \frac{\left| \LF At^n - (t-1)^n\RF_{t=0}\right|^{n-2}}{\left|A-1\right|^{n-2}},\\
= n^n \left|A\right|^{n-1} \left|A-1\right|^{2(1-n)}\,,\label{id4}
\end{aligned}
\ee
\be
\begin{aligned}
\prod\limits_{j=2}^{n} \prod\limits_{k=1}^{n}   \left| t_j^{\infty} - t_k\right|&=
\prod\limits_{k=1}^{n} \frac 1 n \left| t_k^n - (t_k-1)^n\right| 
= n^{-n} \prod\limits_{k=1}^{n} \frac{\left| 1-A\right|}{\left| 1- A^{\frac{1}{n}}e^{\frac{2\pi ik}{n}}\right|^n} \\
&= n^{-n} \left|1-A\right|^n\prod\limits_{k=1}^{n} \left|t_k\right|^n 
%&=  \frac{\left|1-A\right|^n}{\left|1-A\right|^n},\\
=n^{-n}\,,
\end{aligned}
\ee
%%
%\be
%\prod\limits_{j=2}^{n} \left| t_j^{\infty} - t_k\right| = \frac{1}{n}\left|t_k^n - (t_k-1)^n\right|,\label{id6}
%\ee
%%
\be
\begin{aligned}
\prod\limits_{j=2}^n\prod\limits_{k=2, k \neq j }^n \left| t_j^{\infty} - t_k^{\infty}\right| 
&= \prod\limits_{j=2}^n \frac{1}{n} \left| \frac{d}{dt} \LF t^n - (t-1)^n\RF \right|_{t=t_j^\infty}\\
&=\prod\limits_{j=2}^n \left|  \LF t_j^{\infty}\RF^{n-1} - (t_j^\infty-1)^{n-1}\right| \\
&=  \prod\limits_{j=2}^n  \left| 1-e^{\frac{2\pi i(j-1)}{n}}\right|^{2-n}
= \prod\limits_{j=2}^n \left|  t_j^\infty\right|^{n-2} 
= n^{2-n}\,.\label{id8}
\end{aligned}
\ee
Using these identities we can write the expressions  \eqref{eq:bj} and  \eqref{eq:ai} as
\begin{align}
\prod\limits_{i=1}^{n} \left| a_i\right| & = n^{mn} \left| A \right|^{m(n-1)} \left| A-1\right|^{m(2-n)}\,, \\
\prod\limits_{j=1}^{n} \left| b_j\right| &=  n^{-m(n+2)} \left| A-1\right|^{nm}\,, 
\end{align}
and substituting in the correlator gives
\begin{align} \label{eq:resultcorrelator}
\langle [ \tilde  \sigma_m & (\infty)]^n \tilde  \Sigma^{(n)}_{1}(1)  \Sigma^{(n)}_{1+\ell}(x,\bar x) [\sigma_m (0)]^n  \rangle  = 
\left[ m^2 \left| A\right|^{m-1} \left|A-1\right|^2\right]^{-\frac{c}{12}\LF n-\frac{1}{n}\RF}\,. 
\end{align}
%
%where $A= e^{i\frac{(\alpha+2\pi j)}{m}}$. 

\setcounter{equation}{0}
%%%%%%%%%%%%%%%%%%%%%%%%%%%%%%%%%%%%%%%%%%%%%%%%%%%%%%
\section{Dominant contributions to the entwinement of $M=0$ BTZ} \label{SqrtN}

In this appendix, we show that  in a typical state of the zero mass BTZ black hole the sum
\be  \label{eq:sum}
\sum_m m N_m \left| 2m\sin\LF \frac{ \alpha+2\pi \ell}{2m}\RF\right|^{-\frac{c}{6}(n-\frac 1 n)}
\ee
appearing in the correlator \eqref{eq:EWcyl} is dominated by strings with $m \sim O(\sqrt{N})$. 

We assume we are working at large enough $N$, such that deviations from typicality are small. The total number of strings of length $m$ in a typical state is 
\be
N_m = \frac{8}{\sinh \beta m}
\ee
with $\beta \simeq \sqrt{2}\pi/\sqrt{N}$~\cite{Balasubramanian:0508}. Since $m$ is an integer, we can write the total number of strands as 
\begin{equation} \label{eq:deltam}
N = \sum \limits_m m N_m \Delta m\,,
\end{equation}
with $\Delta m=1$, and by redefining $m \equiv xN$, $\Delta m \equiv \Delta x N$:
\begin{equation}
1= N \sum\limits_x x N_x \Delta x.
\end{equation}
In the limit $N \to \infty$ this becomes the integral
\begin{equation}
1= N \int\limits_{0}^{\infty} dx \frac{8x}{\sinh (\sqrt{2}\pi x\sqrt{N})} \,,
\end{equation}
which can be split as
\begin{equation}
1= N \int_{0}^{N^{\gamma-1}} dx \frac{8x}{\sinh (\sqrt{2}\pi x\sqrt{N})}
+N\int_{N^{\gamma-1}}^{N^{\gamma'-1}} dx \frac{8x}{\sinh (\sqrt{2}\pi x\sqrt{N})} 
+N\int\limits_{N^{\gamma'-1}}^{\infty} dx \frac{8x}{\sinh (\sqrt{2}\pi x\sqrt{N})} 
\end{equation}
with $\gamma < 1/2$ and $\gamma' > 1/2$. We now observe that the first and last integral vanish in the large $N$ limit.
\begin{itemize} 
\item For $x \in [0,N^{\gamma-1})$, or $m\in [0,N^{\gamma})$, and $\gamma < 1/2$:
\begin{equation}
N\int_{0}^{N^{\gamma-1}} dx \frac{8x}{\sinh (\sqrt{2}\pi x\sqrt{N})} \sim  \int_{0}^{N^{\gamma-1}} dx \frac{8\sqrt N}{\sqrt 2 \pi} =  \frac{8 N^{\gamma-1/2}}{\sqrt{2}\pi} \to 0\,, \mbox{ as } N \to \infty\,. 
\end{equation}
\item For $x \in [N^{\gamma' -1},\infty)$, or  $m \in [N^{\gamma'},\infty)$, and $\gamma' > 1/2$:
\begin{align}
N\int _{N^{\gamma'-1}}^{\infty} dx \frac{8x}{\sinh (\sqrt{2}\pi x\sqrt{N})} & \sim  N\int _{N^{\gamma'-1}}^{\infty} dx \frac{16 x}{e^{\sqrt 2 \pi x \sqrt N}} \\
& \sim \frac{8\sqrt{2}}{\pi} N^{\gamma'-1/2} e^{-\sqrt{2}\pi N^{\gamma' -1/2}} \to 0\,, \mbox{ as } N \to \infty\,. 
\end{align}
\end{itemize}
It follows that
\be
1 \approx N\int_{N^{\gamma-1}}^{N^{\gamma'-1}} dx \frac{8x}{\sinh (\sqrt{2}\pi x\sqrt{N})} = \frac{1}{N} \sum_{m\sim N^{\gamma}}^{N^{\gamma'}} \frac{8m}{\sinh \beta m} 
\ee
and
\be
N \approx  \sum_{m\sim N^{\gamma}}^{N^{\gamma'}} m N_m
\ee
for any $\gamma<1/2$ and $\gamma'>1/2$, and especially for $\gamma,\gamma'$ arbitrarily close to $1/2$ proving that indeed strings with $m \sim O(\sqrt{N})$ dominate the sum \eqref{eq:deltam}. 

To complete the proof, we observe that for fixed $\alpha$ and $l$
\be
2m\sin\LF \frac{ \alpha+2\pi \ell}{2m}\RF \sim O(1)
\ee
is of the same order regardless of $m$. Therefore
\be
\sum_m m N_m \left|  2m\sin\LF \frac{ \alpha+2\pi \ell}{2m}\RF\right|^{-\frac{c}{6}(n-\frac 1 n)} \sim 
N  \left|  2 \sqrt N \sin\LF \frac{ \alpha+2\pi \ell}{2 \sqrt N }\RF \right|^{-\frac{c}{6}(n-\frac 1 n)}
\ee
for all finite $\ell$ not scaling with $N$.

\setcounter{equation}{0}

%%%%%%%%%%%%%%%%%%%%%%%%%%%%%%%%%%%%%%%%%%%%%%%%%%%%%%%%%%%%%%%%%%%%%%%%%
%%%%%%%%%%%%%%%%%%%%%%%%%%%%%%%%%%%%%%%%%%%%%%%%%%%%%%%%%%%%%%%%%%%%%%%%%

%%%%%%%%%%%%%%%%%%%%%%%%%%%%%%%%%%%%%%%%%%%%%%%%%%%%%%%%%%%%%%%%%%%%%%%%%%
%%%%%%%%%%%%%%%%%%%%%%%%%%%%%%%%%%%%%%%%%%%%%%%%%%%%%%%%%%%%%%%%%%%%%%%%%%

\end{document}